\documentstyle[psfig]{caosp}
\begin{document}
\title{The effects of binarity in driving CP phenomena}
\author{J\'{a}n Budaj}
\institute{\lomnica}
\maketitle
\begin{abstract}
A lot of evidence has recently accumulated that physical
characteristics of CP stars ($\delta m_{1}$, Ca/Fe ratio, Li abundance,
distribution of orbital periods, $v \sin i$, $\Delta (V1-G)$, magnetic
field) depend on orbital period or eccentricity.
Consequently, the "tidal mixing + stabilization" hypothesis has been formulated
to account for such features and the "binarity $\times$ magnetism"
hypothesis of Abt \& Snowden (1973) was reanalyzed.
We conclude that all mentioned characteristics and
properties of CP stars, including even CP star magnetism,
may be affected or governed by binarity.
\end{abstract}

\section{Motivation or ``binarity $\times$ magnetism'' hypothesis}

The backbone of the current research of CP stars is their modern
classification scheme by Preston (1974). He clearly distinguished
a non-magnetic sequence of Am and HgMn stars as well as
a magnetic sequence of Ap and some He-weak stars.
While, at present, we have at hand some
explanation of most of the CP star distinguishing characteristics like
the origin of abundance anomalies and slow
rotation of these stars, this is not the case of their magnetism
and binarity. Both of these phenomena are often referred to as primordial
reasons for CP peculiarity in the corresponding area of the HR diagram.
Nevertheless, it is slow rotation which is generally accepted
as a more direct reason for the CP phenomenon and which is closely related to
the primordial causes, namely, being a result of tidal or magnetic breaking
mechanisms (Abt 1965, 1979, Wolff 1983). The reason for these primordial
causes is however not very clear, and this is just what we
aim to touch in these studies.
The pioneering works of Abt (1961) and Abt \& Snowden (1973)
completed and reanalyzed e.g. by Abt (1965), Floquet (1983), 
Abt \& Levy (1985), Gerbaldi et al.(1985),
Lebedev (1987), Seggewiss (1993), North (1994), North at al. (1998) and many
others revealed that the
frequency of occurrence among binaries is much higher for Am than for Ap stars,
short period orbits are much more frequent in Am than in Ap ones
and the N(SB2)/N(SB1) ratio is higher in Am's than in Ap's.
The reversed appearances of magnetism and binarity
in Am and Ap stars suggest a possible relationship between these
fundamental CP characteristics and we will handle such an idea
as a "binarity $\times$ magnetism" hypothesis.
It was first suggested by Abt \& Snowden (1973).
Guided by such motivations we searched
(e.g. Budaj 1995, 1996, 1997a, 1997b, Budaj et al. 1997, Iliev et al. 1997)
for the various possible imprints of a stellar companion on the CP star.

\section{Imprint 1: orbital periods of Am and Ap binaries}
The orbital period distribution (OPD) of Am's was compared with that of 
corresponding normal binaries. It seems that:\\
-- Am binaries do not have orbital periods less than $\approx 1.2^{d}$, except
for a few stars, which is not the case of normal ones
(also e.g. North et al. 1998).\\
But the most interesting puzzle is:\\
-- a gap in the OPD of Am's between 180 and 800 days \footnote{Abt (1965)
already noticed the dip near 100 days in the OPD of Am's. But, at that 
time, he could only conclude that it "may not be real" and it has not been 
mentioned any longer.}, \\
-- which center is replaced by a peak in the OPD of normal stars (5 of 
them).\\
-- Such a dip (160 -- 600 days) was also found in the OPD of Ap binaries.

\section{Imprint 2: rotation of Am binaries}

Based on the current view on Am stars, one would expect a horizontal dividing
line at about $\approx$ 100~km\,s$^{-1}$ which would separate Am from normal 
stars in a plot of equatorial rotation velocity versus $P_{orb}$.
We analyzed the $v\sin i$ versus $P_{orb}$
dependence for Am binaries and the most interesting results are:\\
-- there is an apparent area of synchronization at the short periods;\\
-- the maximum rotational velocity allowed for an Am star, $V_{max}$, is
not a constant but an increasing function of $P_{orb}$ in the range
$4^{d} <P_{orb}<20^{d}$;\\
-- only for the longer periods ($P_{orb}>20^{d}$), $V_{max}$ acquires a
constant value of about 75 km\,s$^{-1}$;\\

\section{Imprint 3: metallicity and lithium in Am binaries}

We accepted the $\delta m_{1}$ index, Ca/Fe ratio
(ratio of the equivalent width of Ca I 6717 and Fe I 6678 lines)
and Li abundance as the metallicity or peculiarity indicators. The smaller the
$\delta m_{1}$ or Ca/Fe, the larger the metallicity or peculiarity.
It turned out that:\\
-- the $\delta m_{1}$ index decreases with the orbital period at least up to
$P_{orb}\approx 50^{d}$ and probably up to
$P_{orb}\approx 200^{d}$;\\
-- the Ca/Fe ratio exhibits a similar tendency, but it is not well defined due 
to the rather small number of stars;\\
-- the lithium abundance may also decrease with
$P_{orb}$ but, like in the case of Ca/Fe, it is not well defined.\\
The dependence on eccentricity seems even more pronounced because:\\
-- the Ca/Fe ratio decreases with eccentricity;\\
-- the Li abundance decreases with eccentricity;\\
-- $\delta m_{1}$ decreases with eccentricity (mainly in
the interval $10^{d}<P_{orb}< 180^{d}$).\\
Moreover, the $\delta m_{1}$ index exhibits a very interesting
behaviour in the diagram $v\sin i$ versus $P_{orb}$ or
$v\sin i$ versus instantaneous orbital period at periastron:\\
-- the curve of constant metallicity, $M_{const.}$, is a
function of $P_{orb}$ and is parallel to the $V_{max}$ curve!

\section{Imprint 4: the $\lambda 5200$ depression and magnetic fields in
Ap binaries}

We studied the $\Delta (V1-G)$ index which is related to
the $\lambda 5200$ depression and mean effective magnetic field,
$<B_{e}>$ (e.g. Glagolevskij et al. 1986). The larger the $\Delta (V1-G)$
index, the more pronounced the depression.
We have succeeded in disentangling the $e$ and $P_{orb}$ dependences
and discovered that:\\
-- both $\Delta (V1-G)$ and $<B_{e}>$ decrease with eccentricity
if $P_{\rm orb} < 160^{d}$
\footnote{A decrease of $\Delta (V1-G)$ with eccentricity was already 
indicated by Gerbaldi et al. (1985).};\\
-- both parameters also very probably increase with $P_{orb}$ if
$P_{\rm orb} < 160^{d}$, while the trend seems opposite at the larger 
$P_{orb}$ beyond the period gap.\\

\section{Discussion, interpretation and disentangling}
It is hardly possible to explain most of the above-mentioned points
in the framework of current views on CP stars, i.e. as a result of
slow rotation and magnetism in the corresponding area of the HR diagram,
reducing the role of binarity effects just to slowing down the rotation.
In the following we briefly summarize our interpretation of the above points.

We showed that the lower boundary
of about $P_{\rm orb}\approx 1.2^{d}$ in the OPD of Am's corresponded
to the semidetached systems with the Am stars radii $R=3R_{\odot}$.
One cannot, however, exclude that the Am phenomenon is disturbed slightly
before
the Roche lobe is filled in what would then result in a little lower radii 
of Am stars.
Consequently, the Am stars' rotation in this area of synchronization
could be limited by its orbital period (at some fixed stellar
radius) rather than by the generally accepted critical rotation velocity
of about 100 km\,s$^{-1}$.

We argue that a decrease of $\delta m_{1}$, Ca/Fe, or Li abundance
(i.e. increase of peculiarity) with $P_{\rm orb}$ could be explained
by the hypothesis on "tidal mixing + stabilization".
We propose the mechanism of tidal mixing which acts to smooth the
chemical anomalies built by diffusion.
It gets more intensive when the components are closer and weakens
when (pseudo-)synchronization is approached.
The fact that the increase of peculiarity is observed only in a limited range
of orbital periods (up to $P_{orb}\approx 50^{d}$ or $P_{orb}\approx
200^{d}$) gives rise to the idea of some kind of stabilization mechanism
within this range. The latter arises because, without some ``stabilization'',
the peculiarity or chemical anomalies should progressively increase with
$P_{orb}$ up to infinite $P_{orb}$ and, at the same time, approach the status
of ``single'' stars, which is not observed.

The behaviour of the $V_{max}$ curve, which is not a constant but
a function of $P_{orb}$, sheds more light
on the well known overlap in rotation velocities of normal and Am stars
for $40< v < 100$ km\,s$^{-1}$ (Abt \& Hudson 1971) or the deficiency of  
normal
stars for $P_{orb}>2.5^{d}$ (Abt \& Bidelman 1969) because it intersects 
the $R=2.1\, R_{\odot}$ synchronization curve at 
$v=35$ km\,s$^{-1}$ and $P_{orb}=3^{d}$.
The parallel behaviour of the $M_{const.}$ and $V_{max}$ curves is hardly an
accident but it is rather a natural consequence of the idea that each curve
corresponds to different intensity of turbulence in the
$v$ versus $P_{orb}$ diagram.

The fact that there is a peak in normal stars within
the period gap in Am's supports the presence of the mentioned
gap. \footnote{One could estimate its significance to $1/2^{5}$
as there are 5 normal stars. The gap in Ap's is also significant approximately
at the level of $2 \sigma$.}
Thus, three independent samples (Am, Ap, normal A4-F1) indicate
that there might be some breaking point in the
(magneto)hydrodynamics of an AV-type binary at the orbital period
of several hundred days.
We suggest the "tidal mixing + stabilization" hypothesis to account for this
gap and propose that there is a peak in the turbulence within
the period gap so that the He superficial convection zone  cannot 
disappear due to He settling.
This could result in the observed pattern of all three
samples.

Finally, our findings that the degree of peculiarity in Ap's
depend on orbital elements strongly supports the "binarity $\times$ magnetism"
hypothesis and we can conclude that magnetism, as a necessary condition
for Ap phenomena, is (1) affected by binarity or (2) the opposite.
In fact, this second possibility cannot be definitely excluded
based on the above arguments (including those mentioned in Sect. 1) only.
Abt \& Snowden (1973) preferred just this possibility and
suggest e.g. that "for those Ap stars having strong magnetic field, the
formation of binaries with separation $10^{6}-10^{9}$ km is inhibited ...".
Nevertheless, the second possibility seems to be more complicated and thus
more unlikely in the
light of our recent findings, due to a rather complex behaviour of Ap
peculiarity with respect to orbital elements.
Thus, we favor the idea that binarity "governs" the
magnetism. At present it is not very clear what could be the reason of such an
interplay but one can speculate that (pseudo-)synchronization might play
an important role.
Generally, the shorter the orbital period, the higher the degree of
(pseudo-)synchronization and the binary components tend to rotate as rigid
bodies. One can expect that this will suppress the differential stellar
rotation, which is thought to drive the magnetism, at least in the case of the
Sun. It could then result in a deficit of short periods
or an increase of peculiarity with $P_{orb}$
or low frequency of occurrence or low SB2/SB1 ratio in Ap binaries.
Even the decrease of their peculiarity with eccentricity
could be understood because highly eccentric orbits have higher velocities at
periastra than circular ones, thus also a higher degree of
pseudo-synchronization.
Also it would not be very surprising to observe just the
opposite behaviour, i.e. more
pronounced anomalies at larger eccentricities or relative enhancement of short
period orbits (except for $P_{orb}<1.2^{d}$) in non-magnetic Am binaries.
Nevertheless, the latter facts could also be explained outside
the scope of the "binarity $\times$ magnetism" hypothesis as a consequence of 
the mentioned "tidal mixing + stabilization" hypothesis, since tidal mixing
should weaken as pseudo-synchronization approached.

To conclude, there is for the first time reliable observational evidence 
(except for synchronization and circularization) that
the hydrodynamics or magneto-hydrodynamics of a ``detached binary'' component 
is affected by its companion in a surprisingly broad interval of orbital 
elements .


\vspace{-1.0mm}
\acknowledgements
I thank the editors for carefull proofreading the contribution.
This work has been supported by the VEGA Grant No. 4175.

\end{document}